\def\plotone#1{\centering \leavevmode
\epsfxsize=\textwidth \epsfbox{#1}}

\documentstyle[11pt, aas2pp4,psfig]{article}    

\def\sqig{$\sim$}

\def\degrees{$^{\circ}$}

\def\source{XTE\,J1855-026}
\def\src{XTE\,J1855-026}

\begin{document}
\title{
RXTE Observations of the X-ray Pulsar XTE J1855-026
- a Possible New Supergiant System}
\title{\today}

\author{R.H.D. Corbet\altaffilmark{1}, F.E. Marshall,
A.G. Peele\altaffilmark{2}, \& T. Takeshima\altaffilmark{1}}

\affil{Laboratory for High Energy Astrophysics, Code 662,\\
NASA/Goddard Space Flight Center, Greenbelt, MD 20771}
\altaffiltext{1}{Universities Space Research Association}
\altaffiltext{2}{National Research Council Research Associate} 

\begin{abstract}
A new X-ray source, \src, was discovered during RXTE scans along the
galactic plane.  The source shows pulsations at a period of 361 s and
also modulation at a period of 6.1 days which we interpret as the
orbital period of the system. The X-ray spectrum above \sqig3 keV can
be fitted with an absorbed power law model with a high-energy cut-off,
and an iron emission line at approximately 6.4 keV. We interpret these
results as indicating that \src\ is likely to consist of a neutron star
accreting from the wind of an O or B supergiant primary. A less likely
interpretation is that \src\ is instead a Be/neutron star binary,
in which case it would have the shortest known orbital period for such
a system.

\end{abstract}
\keywords{stars: individual (\source) --- stars: neutron ---
X-rays: stars}

\section{Introduction}

High-mass X-ray binaries (HMXBs) are accreting neutron stars or black
holes with an early spectral type (O or B) mass-donating stellar
companion.  These objects lie near the galactic plane and so,
unless unusually close,
typically
are found at low galactic latitudes.  Surveys of the galactic plane have thus
revealed a number of HMXBs, and observations with the Ginga satellite in
particular revealed a large number of sources in the region of the plane
located in Scutum (e.g.  Koyama et al. 1990) which may correspond to
the 5 kpc galactic arm.  Many of these objects are transient and are
thought to be Be/neutron star binary systems. Their transient nature
arises because of both the variable envelope around a Be star, from
which a compact object accretes, and the eccentric orbit of the compact
object as accretion may often occur only close to the time of
periastron passage.  In addition to the transient Be star systems,
other interesting X-ray sources have been found in this region of the
plane such as the ``braking X-ray pulsar'' AXJ1845.0-0300 (Torii et
al.  1998).

In order to investigate this region in more detail and, in particular,
to search for new transient sources, we are monitoring this section of
the galactic plane using the Proportional Counter Array (PCA) on board
the Rossi X-ray Timing Explorer (RXTE) satellite.  We report here the
discovery of a new source, the detection pulsations with a 361 s period,
a measurement of the orbital period with the RXTE
All-Sky Monitor (ASM), and a measurement of the X-ray spectrum
including the detection of an iron line.

\section{Observations and Analysis}

In this paper we present the results of observations of \src\ that have
been made with the instruments onboard RXTE (Bradt, Rothschild, \&
Swank 1983): the Proportional Counter Array (PCA),
the High Energy X-ray Timing Experiment (HEXTE), and the All
Sky-Monitor (ASM).

The PCA is described in detail by Jahoda et al. (1996).  This detector
consists of five nearly identical, Proportional Counter
Units (PCUs) sensitive to X-rays with energies between 2 - 60 keV with
a total effective area of \sqig6500 cm$^2$. The PCUs each have a
multi-anode xenon-filled volume, with a front propane volume which is
primarily used for background rejection.  For the entire PCA across the
complete energy band the Crab produces a count rate of 13,000
counts/s.  The PCA spectral resolution at 6 keV is approximately 18\%
and the field of view is 1\degrees\ full width half maximum.

The high-energy HEXTE instrument on board RXTE (Rothschild et al. 1998)
is coaligned with the PCA and is sensitive to X-rays in the range of 15
- 250 keV.  HEXTE consists of two clusters, each of which itself
consists of four phoswich scintillation detectors. During the
observations of \src\ the two clusters were rocked to positions
$\pm$1.5\degrees offset from the source position every 32 seconds so as
to provide a continuous monitor of the background level.

The All-Sky Monitor (Levine et al. 1996) consists of three similar
Scanning Shadow Cameras, sensitive to X-rays in an energy band of
approximately 2-12 keV, which perform sets of 90 second pointed
observations (``dwells'') so as to cover \sqig80\% of the sky every
\sqig90 minutes.  The analysis presented here makes use of both daily
averaged light curves and light curves from individual dwell data.
Light curves are available in three energy bands: 1.3 to 3.0 keV, 3.0
to 4.8 keV, and 4.8 to 12.2 keV. The Crab produces approximately 75
counts/s in the ASM over the entire energy range and ASM observations
of blank field regions away from the Galactic center suggest that
background subtraction may yield a systematic uncertainty of about 0.1
counts/s (Remillard \& Levine 1997). In order to be able to obtain the
light curve of an X-ray source the location of the source on the sky
must be known to an accuracy of at least a few arc minutes.

Scans parallel to the galactic plane in the Scutum Region have been
made approximately every other week starting 24 February 1998.  Each
observation consists of scans between galactic longitudes 15\degrees
and 40\degrees\ at galactic latitudes of -1.5, 0.0, and 1.5\degrees.
The subject of this paper, XTE J1855-026, is in the field-of-view of
the PCA for about 20 seconds during each scan at a latitude of
-1.5\degrees.

The new X-ray source \src\ was first noted as active on 1998 May 6th.
An additional scan was then made across the source position
perpendicular to the galactic plane to obtain an
improved location for this source.  A further pointed observation was
then made on May 12 from 19:12 to 21:37. Data were collected in the
``Good Xenon'' mode which provides maximum spectral and timing
information from the PCA. However, for simplicity of analysis, in this
paper we primarily utilize ``Standard Mode 2'' data collected at the
same time.  This provides the same spectral information as the Good
Xenon modes although at a reduced time resolution of 16 seconds. This
is sufficient for our analysis presented here due to the relatively
long source pulsation period. Data were collected from
HEXTE in ``E\_8US\_256\_DX1F'' mode as well 
as the usual archive mode but 
our analysis here uses only the archive mode data. 

\section{Results}

\subsection{Source Location}




The position of \src\ was determined by comparing the observed
light curve in the 2-10 keV band during scans across the source 
with the expected
behavior as the source moved through the PCA field-of-view.
The galactic longitude was determined from the scan at a latitude
of -1.5\degrees\ on 6 May 1998, and
the galactic latitude was determined using
a scan on 8 May 1998 at the longitude estimated from
the 6 May scan.
In determining the source position we also included a model of
emission from the nearby source X1845-024. Also present,
but not included in the model, is unresolved emission
from the galactic ridge, which is difficult to treat
because of the non-imaging nature of the PCA.
From these scans we derive a best-fit position
of RA = 18h 55m 30s, Dec. -02\degrees\ 34\arcmin\ 52\arcsec\ with
an estimated uncertainty at the 90\% confidence level of 2 arc minutes including
systematic effects.

We have 11 scanning observations of the Scutum region obtained
between 1998 May 11 and September 23.
The 2-10 keV flux detected
from the source during scans varied from a lower
level of an upper limit of
about 10 counts/s to a maximum of 136 $\pm$ 15 counts/s on 6 May 1998.





\subsection{Pulsations}
The light curve obtained from the pointed PCA observation is shown in
Figure 1 and directly from the light curve itself pulsations can
clearly be seen. This is confirmed by a Fourier Transform of the PCA
light curve which gives the power spectrum shown in Figure 2.
There is strong modulation at a period of \sqig360 s. A sine wave fit to
the light curve yields a pulse period of 361.1 $\pm$ 0.4 s.  The light
curve folded on that period in several energy bands is plotted in
Figure 3.  We note that the pulse profile shows little change between
the different energy bands.

\subsection{Spectrum}
A spectrum of \src\ was extracted using layer 1 data from the PCA, as
the signal to noise ratio is highest for this layer for faint
sources.  An instrumental response matrix was generated using version
2.36 of ``pcarsp'' which accounts for small time dependent gain changes
in the PCA. An
estimate of the time-dependent background flux was obtained using
``pcabackest'' together with a
model suitable for faint sources. No additional ``systematic'' errors
were added to the resulting spectrum, although the background
estimation software itself does somewhat overestimate the
errors (K. Jahoda private communication).
A HEXTE spectrum were extracted for each of the clusters
using standard software
which includes background subtraction via the off-source rocking
position
observations. Due to relative calibration uncertainties, in all spectral
fits the normalization of the spectrum was a free parameter
for both the PCA and HEXTE spectra with all other parameters
linked.
We first fitted the combined PCA/HEXTE spectrum
with the standard pulsar model (e.g.  White, Swank, \& Holt 1983) of an
absorbed power law with a high energy cut-off. This failed to give a
good fit with large residuals appearing near 6.4 keV. The addition of a
Gaussian line (equivalent width \sqig 360 eV) significantly improved
the fit.  However, a large excess over the model still appears at low
energies (below \sqig 3 keV). Since the efficiency of the PCA falls off
rapidly at low energies, and interstellar absorption strongly reduces
the flux at these energies, any interpretation of this excess as
arising from the intrinsic spectrum of the source must be treated with caution.
Inaccuracies in the wings of the instrument response function might
have major effects for this type of spectrum (K. Jahoda private
communication).  The parameters
of this fit are given in Table 1 and the spectrum together
with the power law plus high-energy cut-off model are shown in Figure 4.

For an unabsorbed spectrum we find a flux from the fitted model of
1.9$\times$10$^{-10}$ ergs cm$^{-2}$ s$^{-1}$ between  2 to 10 keV
while the absorbed spectrum has a flux of 1.1$\times$10$^{-10}$ ergs
cm$^{-2}$ s$^{-1}$.

\subsection{Orbital Modulation in the ASM}

The position obtained from the PCA scans enabled a 2 - 12 keV light curve to
be obtained for \src\ with the ASM, and this is shown in Figure 5.
The mean flux is 0.51 counts per second which corresponds to
approximately 7 mCrab. A Fourier Transform of this light curve yields
the power spectrum shown in Figure 6. A prominent feature of the
power spectrum is a highly significant peak at approximately 6.1 days
and a harmonic of this. To further quantify the modulation on this
period we fitted a sine wave and find a best fit period of 6.067 $\pm$
0.004 days and an amplitude of 0.12 $\pm$ 0.02 ASM counts/second.  In
Figure 7 we show the ASM light curve folded on this period.
The epoch of maximum flux derived from the sine wave fit
is MJD 50289.1 $\pm$ 0.15 and the initial source detection
and the pointed observations thus both occurred at orbital phases of
approximately 0.25.



In order to investigate any long term source variability we divided the ASM
light curve into three sections and examined these for any change in
the orbital modulation amplitude or mean intensity and no significant
changes were found, although the last data section is noisier due to
the inclusion of observations performed when the source was close to
the position of the Sun.

\section{Discussion - the Nature of \src}

Many of the objects found during slews of this region are transient and
are suspected to belong to the large Be/neutron star class of high mass
X-ray binary. However, the orbital period of 6 days would be very short for a
Be star system. The shortest orbital period known for such systems is
16.7 days and belongs to the extreme system A0538-66 (e.g. Skinner et
al. 1980). A search for cataloged objects within this
region using Skyview and Simbad did not find any objects within
our position for \src. Although an earlier preliminary position
for the source had included  
the 12th
magnitude Be star BD -02\degrees4786 at RA = 18 56 03.36, dec. = -02 37
31.1 (Marshall \& Takeshima 1998) this is now apparently
excluded by the more complete position determination presented here
(Section 3.1).

Although there are a few X-ray pulsars that are
found in low-mass systems, the
vast majority of X-ray pulsars are found in HMXBs (e.g.
Bildsten et al. 1997).
The known high-mass X-ray binaries show a clustering into three groups
(Be stars, Roche-lobe overflow, and supergiant stellar wind accretion)
when plotted on a diagram of pulse period against orbital period
(Corbet 1986).  The Be star systems show a general correlation
between these two parameters. If the parameters of \src\ are plotted on
such a diagram (Figure 8) then the source lies in the middle of the
region that, so far, has only been occupied by supergiant wind
accretion driven systems.

Supergiant wind accretion systems, although exhibiting significant
short-term variability (see e.g. Haberl, White \& Kallman 1989)
probably caused by inhomogeneities in the stellar wind, would be
expected to be essentially persistent sources integrated over longer
time scales.  They would not be expected to show the on/off type
behavior exhibited by Be star systems which occurs due to the
appearance and disappearance of their circumstellar envelopes.  The
slews over \src\ with the PCA do show source variability but the ASM
light curve indicates persistence during the period observed. We note
that some of the variability from the slew observations may be
due to the short slew time across the source (20 s) compared to
the 361 s pulse period.  If
\src\ is persistent then it should also have been detected in other
observations that covered this region.  Ginga undertook several scans
of this area (Koyama et al.  1989, 1990) and \src\ lies close to the
nominal error box for source number 6 of Koyama et al. (1990) (= source
number 1 of Koyama et al. 1989).  In addition, the EXOSAT slew catalog
source EMS B1851-024 (Reynolds et al.  1998) has an error box which
lies close to that of \src.  If these sources detected with Ginga and
EXOSAT do indeed correspond to \src, then this would be additional
evidence, beyond that provided by the multi-year ASM lightcurve, that
\src\ is a persistent source.

Further evidence for the supergiant nature of \src\ may come from the
orbital light curve and the X-ray spectrum.  The shape of the ASM light
curve folded on the orbital period is somewhat reminiscent of the
supergiant systems Vela X-1 and 2S0114+650 (e.g. Corbet, Finley \& Peele
1999).  The spectral characteristics of supergiant and Be star systems
are also different and supergiant systems typically show much stronger iron
line emission which can also be highly orbital phase dependent (e.g.
Ohashi et al. 1984, Nagase et al. 1994). The strength of the iron line
detected from \src\ is consistent with a supergiant classification. The
iron line parameters appear to indicate a significant width (0.4 keV)
although, with the spectral resolution of the RXTE PCA, we cannot
determine if this is actually due to the presence of multiple line
components. Although it is likely that the apparent low energy excess
in \src\ is an instrumental artifact we note that, for example, the supergiant
system Vela X-1 does show strong low energy emission lines from neon,
magnesium and silicon  (Nagase et al. 1994) and similar emission from
\src\ might contribute to a low energy excess over a simple model.

Many of the transient sources detected with Ginga in this region by
Koyama et al. (1989, 1990) are suspected to lie in the ``5 kpc arm"
which is at a distance from us of approximately 10 kpc. For this
distance the unabsorbed flux measured with the PCA corresponds to a
luminosity of 2$\times$10$^{36}$ ergs s$^{-1}$ which is broadly
consistent with the flux expected from a wind accretion system
(e.g. van Paradijs 1995).
The
absorption that we measure is also comparable to that found for other
sources in this region (Koyama et al. 1989, 1990) and is
consistent with the predicted interstellar absorption
for this region (Hayakawa et al. 1977).
We note that the X-ray absorption, if
interstellar rather than due to material local to
the neutron star, would imply a very large optical reddening of
approximately E(B - V) = 24 (Bohlin, Savage \& Drake 1978) greatly
impeding observations at other wavelengths such as optical and
in soft X-rays. Such an interstellar absorption would also
indicate that BD -02\degrees4786 is not the optical counterpart.




\section{Conclusion}
The orbital period and pulse period of \src\ suggest that this may be a
supergiant wind accretion powered high mass X-ray binary. As members of
this class show persistent emission, most readily observable galactic
members of this class may already have been found and this would be the
first new member of this class discovered for some time.  However, if
\src\ is, instead, a Be type system, then it would have the shortest
known orbital period of all members of this class. Further
observations to measure the mass function via Doppler shifts of the
pulses and determine a more precise position would be valuable.

\acknowledgments
This paper made use of quick look data provided by the RXTE ASM team at
MIT and GSFC. We thank many colleagues in the RXTE team for useful
discussions on many aspects of RXTE data analysis.
This paper made use of the SIMBAD data base maintained
by the Centre de Donn\'es astronomique
de Strasbourg.

\pagebreak
\noindent
{\large\bf Figure Captions}

\figcaption[pca_lc.ps]
{PCA background subtracted light curve of \src. The start time corresponds to 1998 May 12, 19:12 UT.}

\figcaption[pca_ft.ps]
{Power spectrum of the PCA light curve of \src.
The strongest peak is at a period of approximately 361s. Aliasing
caused by the gap in observations due to Earth occultation is also
present together with a harmonic of the 361s period and its own alias
pattern.
}


\figcaption[multi_fold.ps]
{PCA background subtracted
light curve of \source\ folded on the pulse period
in several energy bands and also summed over the entire 2 to
40 keV energy band (top panel).}

\figcaption[xspec.ps]
{PCA and HEXTE spectra of \src\ and residuals from an absorbed power
law, 6.4 keV emission line and high-energy cut-off. Parameters of the
spectral fit are given in Table 1.
}

\figcaption[asm_lc.ps]
{ASM light curve of \source. Data points are a rebinned and smoothed
version of the standard one day light curves. }

\figcaption[asm_ft_plot.ps]
{Power spectrum of the ASM light curve of \source.}

\figcaption[orbital_fold.ps]
{ASM light curve of \source\ folded on the proposed orbital period.
For clarity, two cycles are plotted. Phase 0 corresponds
to the peak of the fitted
sine curve which occurs at MJD 50289.1 $\pm$ 0.15.
}

\figcaption[stars.ps]
{Pulse period plotted against orbital period for various type of
high-mass X-ray pulsator (cf. Corbet 1986). ``B'' indicates likely Be
star sources, ``R'' sources which are probably powered by Roche-lobe
overflow, and ``W'' shows sources accreting from the wind of a
supergiant star. ``O'' marks the properties of OAO 1657-415 which has
some properties of both wind accretion and Roche-lobe overflow sources.
The parameters of \src\ itself are marked by the circled dot.  Source
parameters can be found in Bildsten et al. (1997) supplemented
by Corbet \& Peele (1997),
Corbet, Peele, \& Remillard (1997),
Corbet et al. (1999),
Israel et al. (1998), and
Soffita et al. (1998).
The horizontal bar indicates the range of allowed orbital
periods found for GRO J1948+32 (Chakrabarty et al. 1995)
which is thought to be a Be star system.
}


\begin{table}
\caption{Spectral Fits to RXTE Observations of \src}
\begin{center}
\begin{tabular}{lc}
Parameter & Cutoff Powerlaw \\
\tableline
N$_H$ (10$^{22}$ cm$^{-2}$) & 14.7 $\pm$ 0.6\\
$\Gamma$ & 1.23 $\pm$ 0.03 \\
A$_{PCA}$ & 0.021 $\pm$ 0.002  \\
A$_{HEXTE}$ & 0.013 $\pm$ 0.001  \\
E$_{Fe}$ (keV) &  6.52 $\pm$ 0.03\\
$\sigma_{Fe}$ (keV) & 0.44 $\pm$ 0.06 \\
EW$_{Fe}$ (eV) & 347 \\
E$_{cut}$ (keV) & 14.7 $\pm$ 0.5 \\
E$_{fold}$ (keV) & 27 $\pm$ 2  \\
$\chi^{2}_{\nu}$ (d.o.f.) & 1.06 (125)\\
\tableline
\end{tabular}
\end{center}
All errors are 1$\sigma$ single-parameter confidence levels.
An energy range of 3.5 to 40 keV was used for the PCA spectrum
and 15 to 100 keV for the two HEXTE spectra.

\end{table}

\begin{figure}
\plotone{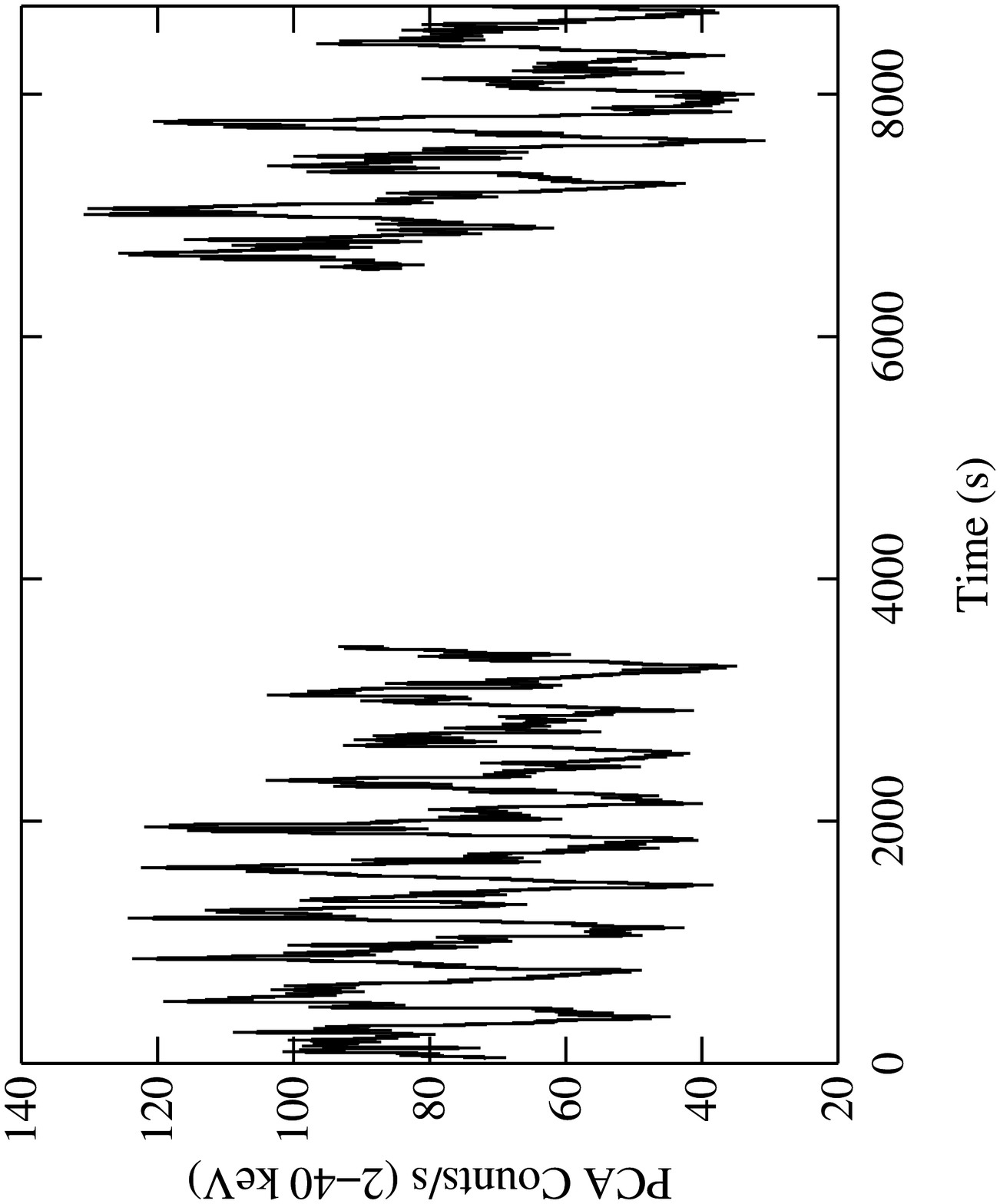}
\end{figure}

\newpage
\begin{figure}
\plotone{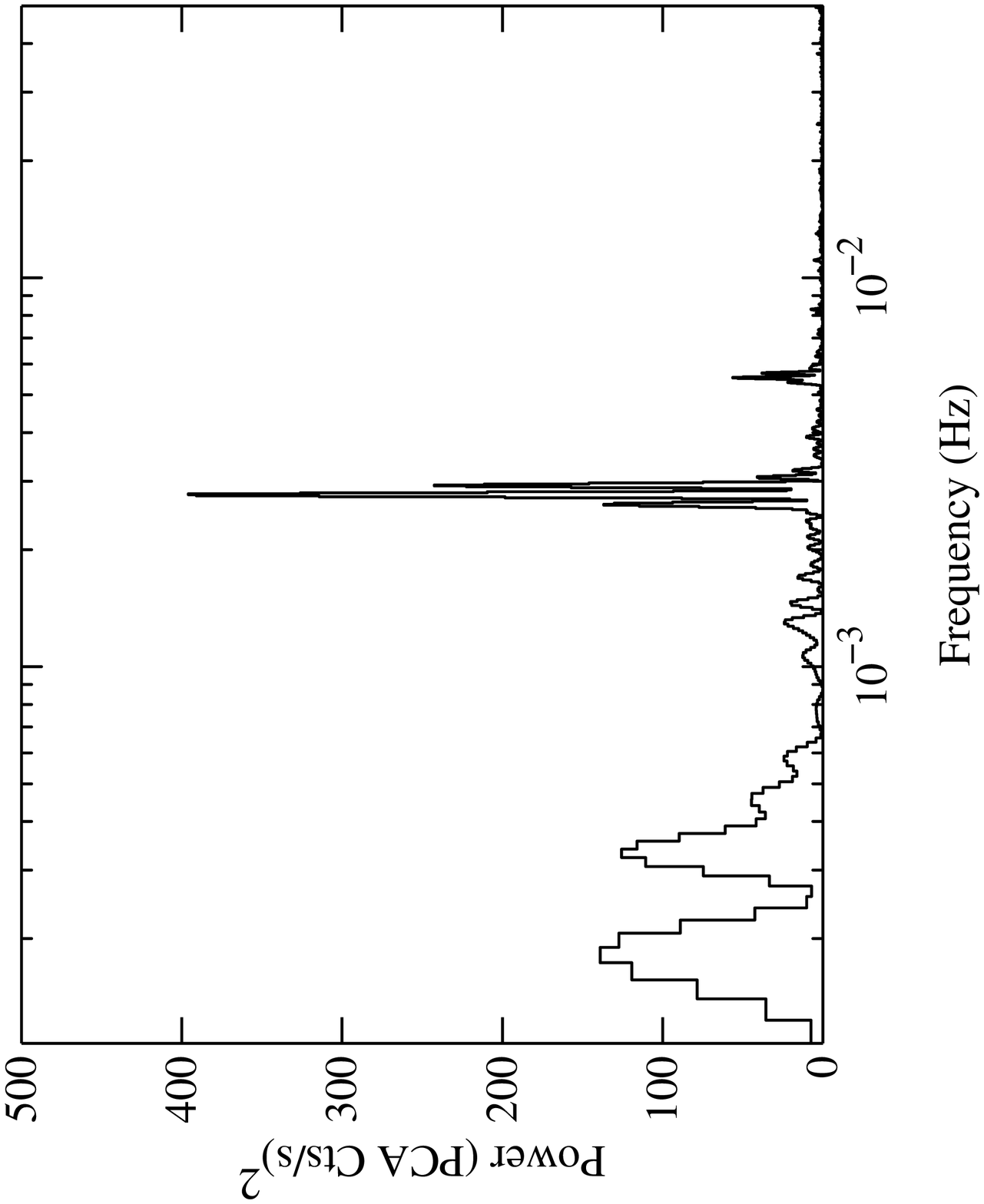}
\end{figure}

\newpage
\begin{figure}
\hbox{
\plotone{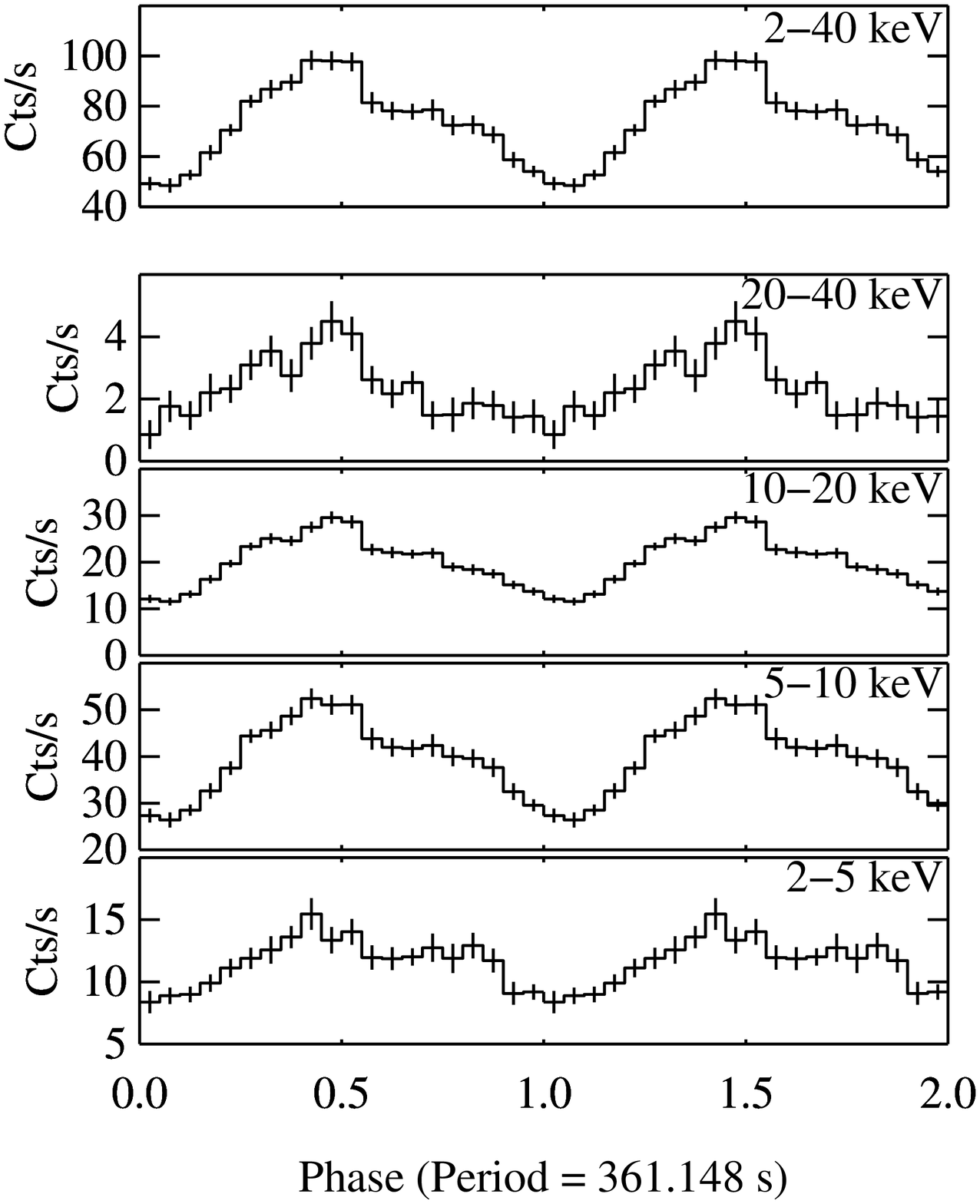}
}
\end{figure}

\newpage
\begin{figure}
\hbox{
\psfig{figure=f4.ps,height=12cm}
}
\end{figure}

\newpage
\begin{figure}
\hbox{
\plotone{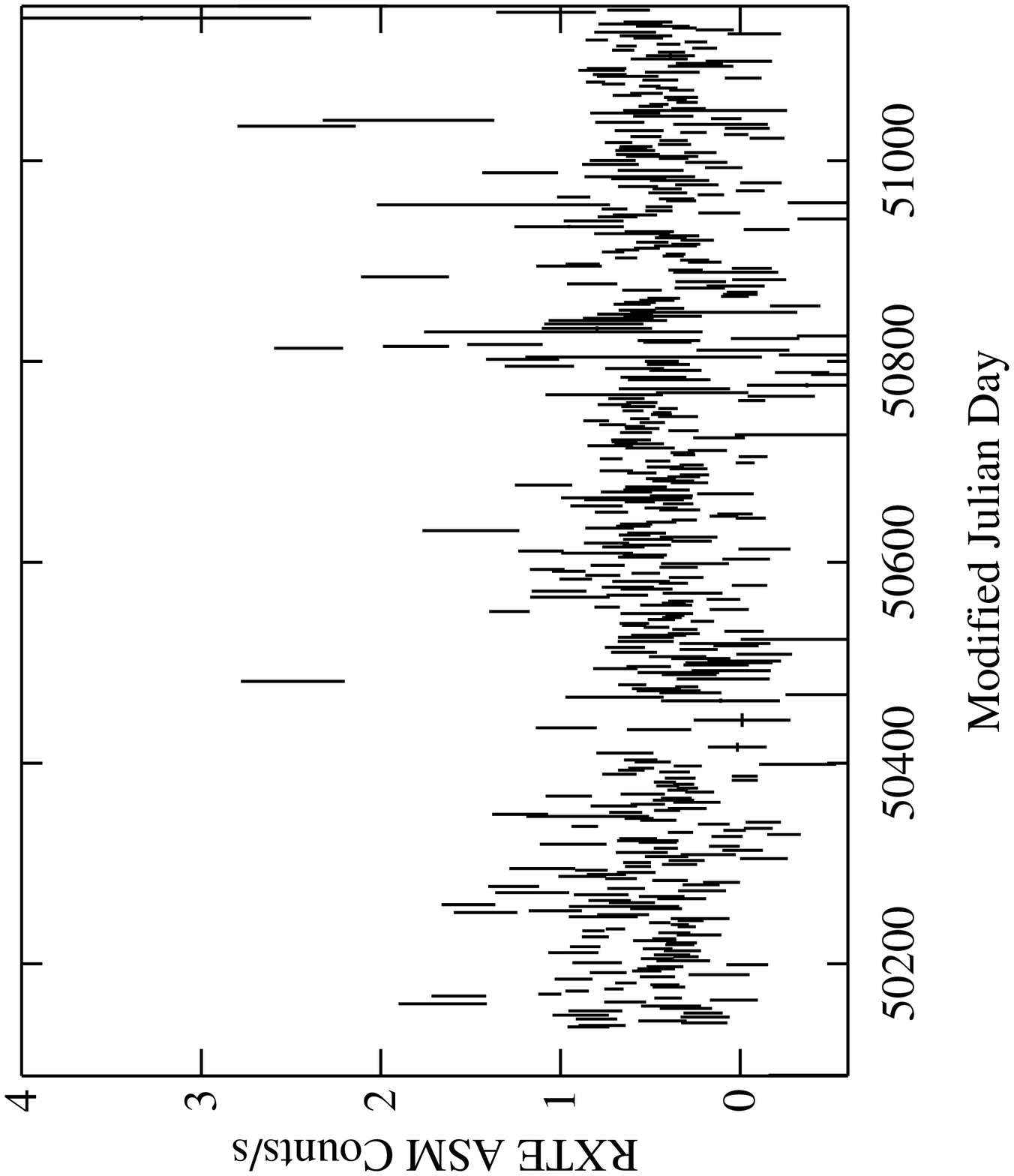}
}
\end{figure}

\newpage
\begin{figure}
\hbox{
\plotone{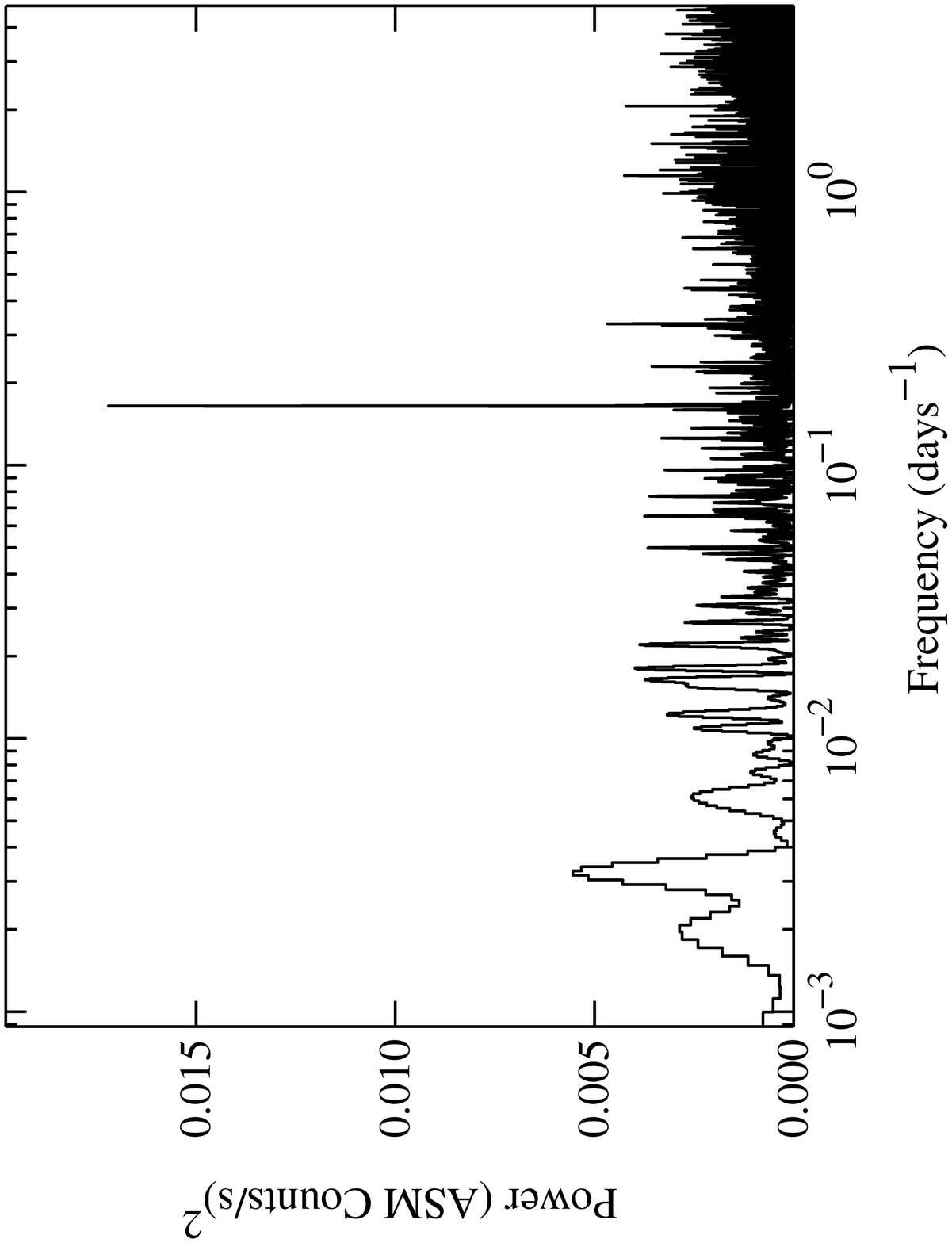}
}
\end{figure}

\newpage
\begin{figure}
\hbox{
\plotone{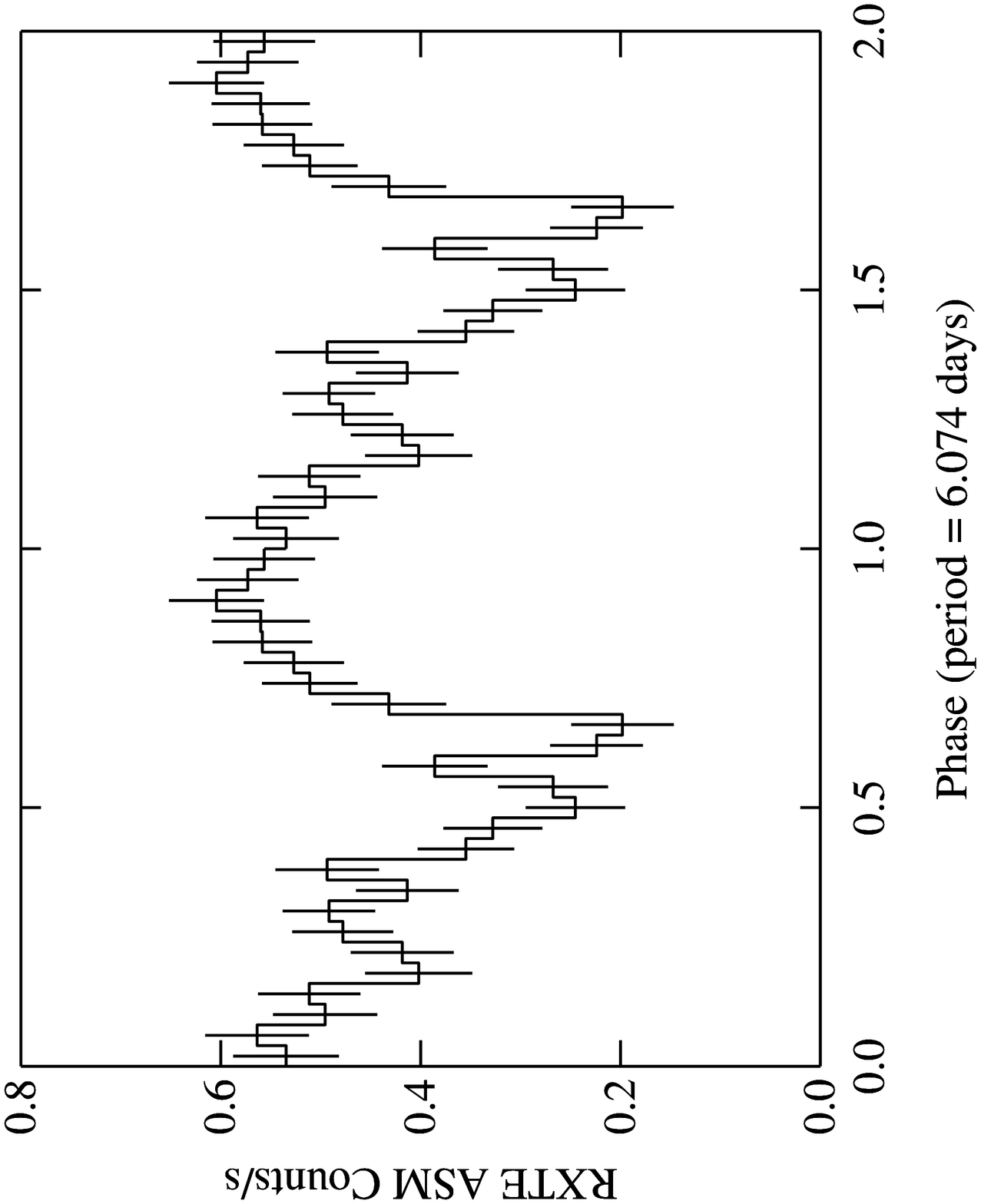}
}
\end{figure}

\newpage
\begin{figure}
\hbox{
\plotone{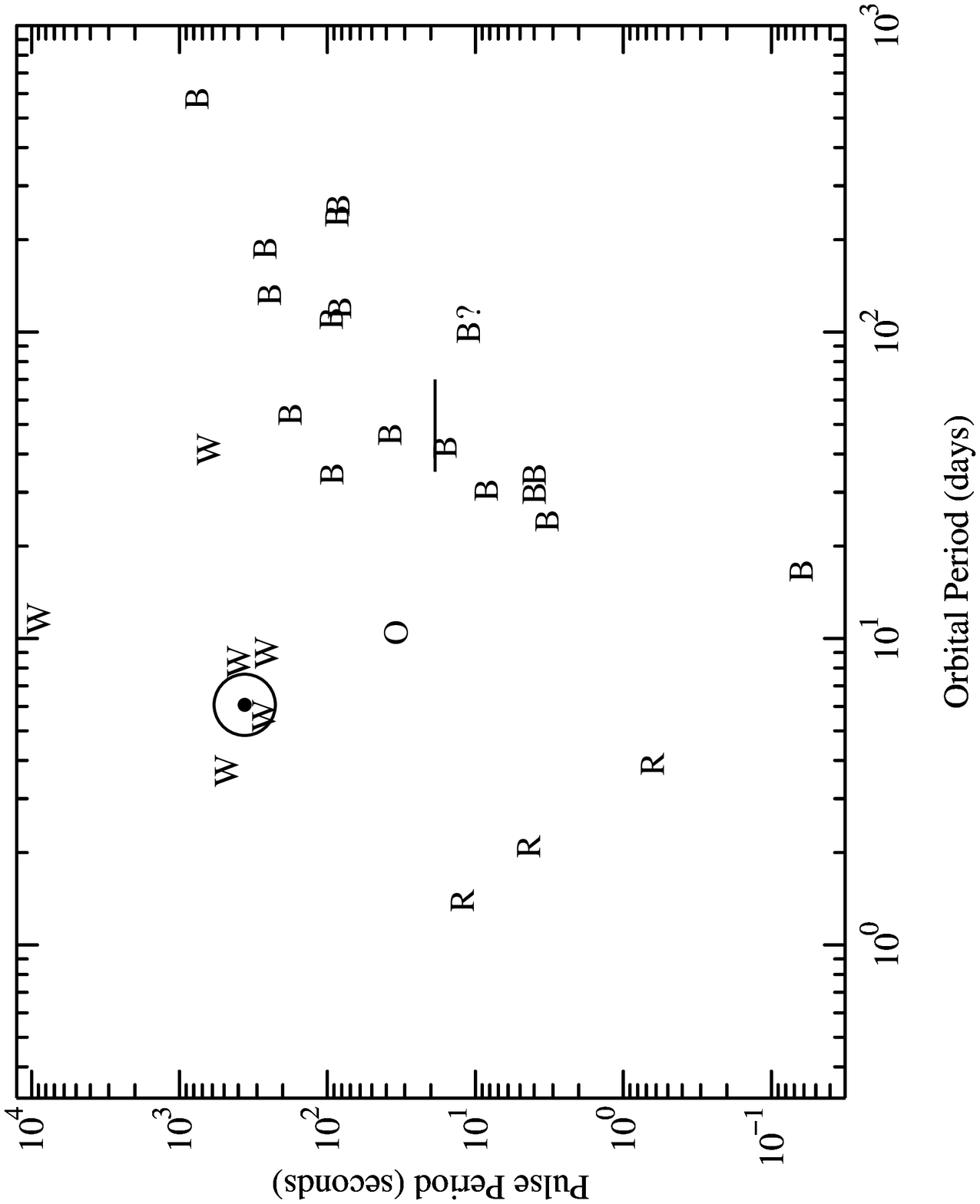}
}
\end{figure}

\end{document}